# ENERGY EFFICIENT MAC PROTOCOLS FOR WIRELESS SENSOR NETWORK: A SURVEY

Eleazar Chukwuka[1] and Kamran Arshad[2]

School of Engineering, University of Greenwich Chatham Maritime ME4 4TB United Kingdom
[1]ce632@gre.ac.uk  [2]k.arshad@gre.ac.uk

*ABSTRACT*

*Wireless Sensor Network (WSN) is an attractive choice for a variety of applications as no wired infrastructure is needed. Other wireless networks are not as energy constrained as WSNs, because they may be plugged into the mains supply or equipped with batteries that are rechargeable and replaceable. Among others, one of the main sources of energy depletion in WSN is communications controlled by the Medium Access Control (MAC) protocols. An extensive survey of energy efficient MAC protocols is presented in this article. We categorise WSN MAC protocols in the following categories: controlled access (CA), random access (RA), slotted protocols (SP) and hybrid protocols (HP). We further discuss how energy efficient MAC protocols have developed from fixed sleep/wake cycles through adaptive to dynamic cycles, thus becoming more responsive to traffic load variations. Finally we present open research questions on MAC layer design for WSNs in terms of energy efficiency.*

## 1. INTRODUCTION

Wireless sensor network is considered to be one of the most influential technologies of the current century [1]. WSN materialised due to the progresses made in micro-electromechanical systems (MEMS) [2] [3] [4] which combines advanced communications and signal processing capabilities [5]. Consequently this led to the production of power constrained low cost tiny sensor nodes [1] [6]. These tiny sensor nodes have capabilities to sense, process and communicate with a remote user through a gateway called the sink. WSN supports ubiquitous computing which is the third generation computer evolution [7]. However, its capabilities, though great, are limited as a result of energy constrains.

Composition of a WSN is made up of the environment that is to be monitored, sensor nodes that are spatially and randomly deployed and the sink node, which is the main interface between the nodes and the user, as shown in Figure1

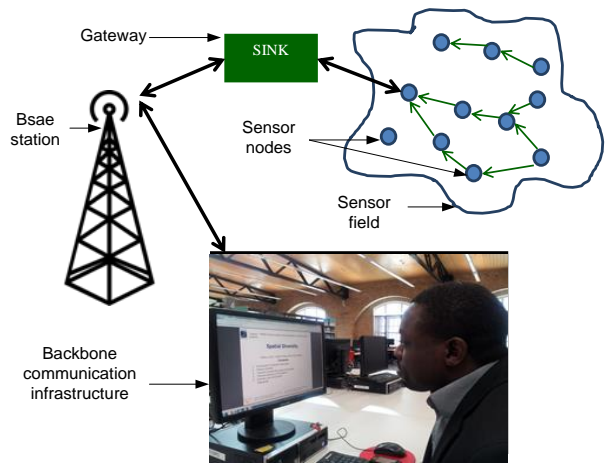

Figure 1: Wireless sensor network architecture

WSN can be deployed for many applications ranging from agricultural [8][9], environmental [10], healthcare delivery [11], military [12], security [13], surveillance [14] [15], home automation [16] and so forth. Smart home concept is also based on WSNs while sensor nodes can be implanted in the human body. These require nodes with good longevity which depends on the source of power supply (usually batteries) among others [17]. These batteries are non-rechargeable due to the nature of deployment and the sensing environment. Hence, it might possibly not be expedient or pragmatic to change the batteries [18]. Furthermore, difficult and adverse terrains like long bridges and high rise civil structures will pose risk to human monitoring (e.g. see Figure 2).

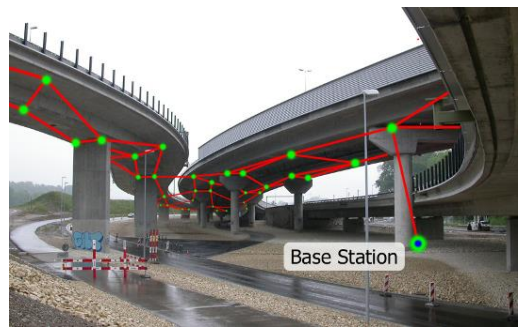

Figure 2: Nodes monitoring a bridge [19]

Enormous research has gone into designing and reviewing energy efficient MAC protocols, reflecting the importance attached to the development of WSNs. Contention-based and schedule-based MAC protocols appear to be the most popular categorisation used in the reviews [20] [21] [22]. In [21] first category is the contention access in which nodes compete to access the channel randomly, and conflicts are resolved by distributed conflict resolution technique [23] [24] [25]. The second is contention free access which offers controlled access to the channel. Nodes only gain access to the medium based on a defined schedule [18] [26]. However the above categorisation is not application specific. A classification according to application was used in [27] which categorised protocols according to the specific problem they aim to solve. But [28] improved on that by focusing on mission critical assignments. Vital performance objectives were used to categorise the protocols into two; data transport delay and reliability, while energy efficiency was put in the back burner. The authors of [29] considered both energy efficiency and data delivery. They separated MAC protocols into four groups; asynchronous, synchronous, frame slotted and multichannel protocols. Synchronous protocols contend with reducing delay and improving throughput. Asynchronous protocols battle with establishing proficient communication between nodes with dissimilar sleep and active plans. Frame slotted protocols assign time slots to nodes to enhance throughput. While this addressed the issue of collision, the challenge is channel utilization in low traffic network. Multichannel mechanism enhances capacity of the network but cross channel communication together with distributed channel assignment is an issue.

These reviews are either application specific or according to the challenges addressed. In this article we show how MAC protocols have been developed to respond to network demand fluctuations. Evolution of MAC protocols is outlined to show the shift from just energy efficiency to efficient utilisation of all resources (energy and bandwidth) together with efficient data delivery. Most of the protocols use duty cycles and contention window to preserve energy. In the early designs, both duty cycle and contention windows were fixed. This survey paper details how, adaptive and subsequently, dynamic duty cycles are being employed in response to network demand fluctuations, enhanced data delivery and channel utilisation thereby conserving energy. Method of medium access was used to group the protocols into four categories, which are controlled access, random access, slotted and hybrid protocols. This categorisation highlights not just the challenges addressed by each group but also indicates the inroads made and how dynamically WSN has been more responsive to network demands.

The rest of this article is organised as follows; a summary of sources of energy inefficiency in WSNs is presented in section 2. MAC protocol classifications and why WSNs differs from traditional wireless networks are detailed in section 3. A detailed survey of existing State-Of-The-Art (SOTA) literature is presented in section 4. Challenges to WSNs in terms of energy efficient MAC are summarised in section 5 while section 6 concludes this article.

## 2. BACKGROUND

WSN is a network of sensor nodes that are randomly and spatially deployed to monitor a phenomenon. Nodes need a source of power supply for them to function; this is usually provided by a battery. The battery might not be replaceable or rechargeable owing to the hazardous terrain in which WSN are normally deployed. Constrained by the power supply, nodes have to optimise their energy consumption but first sources of energy inefficiencies have to be identified. In this section, these sources are identified with radio communication being identified as the most power hungry process. MAC layer controls communication among nodes therefore, we classify MAC protocols used in WSN and show why MAC protocols used in traditional wireless networks cannot be used in WSN.

### 2.1. Sources of energy inefficiency in WSN

Sources of energy inefficiency in WSNs came as a result of different operations of sensor nodes. Events have to be detected therefore sensing [30] [31] [32] is one of the main sources of energy depletion. Another is routing which determines how the sensed data is relayed back to the sink [33] [34]. Processing of sensed data is also an energy draining process in WSNs. Transmitting all raw data consumes considerable energy as well as transmission bandwidth and hence not ideal. To preserve energy, processing of information in the nodes was introduced in [32] [35].

### 2.1.1. Radio Communications

Communication is a major source of power drain in WSNs [27] that involves transmission and reception of data packets. Keeping the radio *on* consumes power while keeping the radio on the sleep mode saves energy but increases latency and offers low spectrum utilisation [36] [37]. WSNs operate in the ISM (Industrial Scientific and Medical) band which is being shared by other wireless devices hence prone to interference and energy wastage.

This survey will focus on energy efficient MAC protocols because communication is the biggest source of energy inefficiency in sensor networks [27] and any reduction in its energy budget will significantly enhance the network lifetime of WSNs. The major sources of energy depletion during communication are transmission with highest drain, followed by receiving and idle modes [38] in that order. The sleep mode has the least energy demand as can be seen from Figure 3 with data from [38]

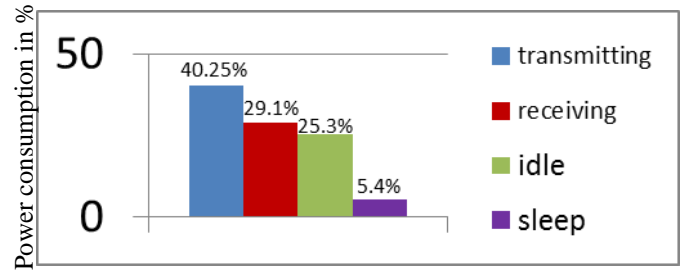

Figure 3: Average power consumption of different node states (modes)

Therefore any protocol that is able to keep transmitting, receiving and idle states to the barest minimum will greatly enhance the network lifetime of WSN. Collision, idle listening, overhearing and control overheads are all causes of energy inefficiencies in MAC protocols [39] which should be reduced.

## 2.2. MAC Protocols

There are different MAC protocols used in the conventional wireless network. MAC protocols used in wired environment cannot be used in wireless environment because collision occurring at the receiver is to be avoided. In wired network the sender detects collision but since signal strength is virtually the same throughout the wired medium, this does not pose any significant problem. However in wireless network the signal strength depreciates in inverse proportion to the square of the distance according Friis free space equation. Another issue is "hidden" terminals, which is as a result of a node "Green" being within transmission distance of nodes "Blue" and "Red" Figure 4, which are not within the radio distance of each other.

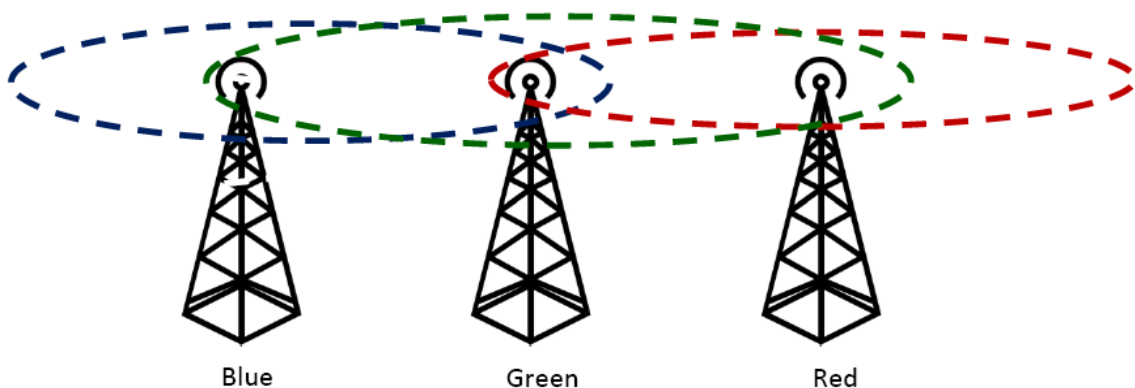

Figure 4: Hidden terminal

Differences between WSN and conventional wireless networks are decentralised control and very long run time in WSNs [27]. Hence ALOHA will not be suitable in WSNs because of collisions [23] [40] and the attendant power drain. CDMA will involve multifarious computing while FDMA require complicated hardware [21], SDMA is usually used in conjunction with other algorithms [41] so does not stand alone. Obviously most of these protocols were not designed with energy constraints as their main concern [42], this is because conventional wireless networks are not as energy constrained as WSNs.

In this article, we group different MAC protocols into four main categories; Controlled Access (CA), Random Access (RA), Slotted Protocols (SP) and Hybrid Protocols (HP). Other categories like low power listening, FDMA, CDMA, beacon enabled etc exist, but this paper will focus on the above categorisations. Next section will present a review of the previous work done.

## 3. SURVEY OF EXISTING STATE-OF-THE-ART (SOTA)

In this section, we highlight current SOTA related with energy efficient MAC protocols for WSN. A number of MAC protocols have been developed for WSNs to address the challenges mentioned in section 2. Earlier protocols were energy efficiency biased but recent techniques in addition, also address others issues like latency, throughput and spectrum utilisation. However no single protocol has been able to address all the sources of power inefficiencies in the radio communication. As an outcome of the survey, we noticed that most of the progresses were made at the expense of throughput, bandwidth utilisation or increased latency. A protocol that can extend the network lifetime of WSN with a less or no trade off will be of great benefit to the society in view of the numerous applications of WSN. Through this categorisation, we show the evolution of MAC protocol from fixed cycles to dynamic cycles as one of the contributions of this article. We first review CA protocols followed by RA then SP and finally HP.

### 3.1. Controlled Access Protocols.

In CA protocols, nodes are allocated time slots using TDMA or FDMA in combination with TDMA. In each time slot a node has access to the shared medium and can transmit without collision. As can be seen in Figure 5, in TDMA, nodes are allocated different times such that at time $t_2$, node $N_3$ has access to the medium. Receiver nodes are synchronised with their sender nodes to wake up at the same time [43]. This protocol enhances energy efficiency by avoiding collision and overhearing. However a lot of overhead is incurred in synchronisation, which together with clock drift is an issue with this protocol [44] [45].

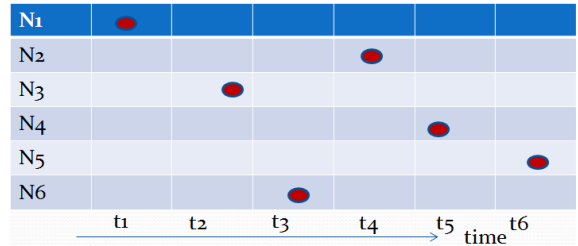

Figure 5: Controlled access protocol

### 3.1.1. Review of Controlled Access Protocols

CA protocols are energy efficient because they avoid collisions which are common with RA protocols (section 3.2). But it is not as scalable and adaptive to time slot assignments and fluctuations associated with WSNs as the later. Also, it is prone to idle listening under low traffic which wastes energy. A protocol, Latency MAC (LMAC) [46], was introduced to reduce idle listening through increasing data arrival prediction accuracy, by increasing sleep cycle which keeps nodes in low power sleep state for longer periods. But this leads to low throughput, high latency and low channel utilisation. For light traffic networks, LMAC offers good energy savings with long sleep cycles. If data arrival rate is predictable, and transmissions occur at a fixed rate, authors of [47] offer an improvement by using time schedule to turn on the radios.

For an event occurring every 10s, the radios are scheduled to come on every 10s, stay on for the duration of the communication and then go back to sleep. This approach avoids idle listening by ensuring that nodes remain in power saving sleep mode, when no transmission is required, which saves energy. However, energy is wasted if there are no transmissions in any interval. This principle is not so efficient since it leads to unnecessary delay in latency or loss of packets due to randomness of data arrival rate. By incorporating adaptive TDMA protocol, [48] [49] enhances the energy efficiency of [47], by making nodes with no packets to transmit to go back to sleep without waiting for its entire slot duration to elapse. Instead of fixed, an adaptive sleep/wake schedule was employed [49]. As the node traffic changes, the schedule adaptively changes according to the traffic. Nodes with light traffic sleep more while nodes with heavy traffic have low duty cycle. If a node has more packets than can be transmitted in one duty cycle, the duty cycle will be extended to accommodate the entire packets, and reversed when packets are less. Thus improving on latency, throughput and channel utilisation, but incurring overhead. By varying the duty cycle according the network need, energy that would have been wasted with fixed cycles which do not consider whether there is traffic or not is saved.

Another protocol that is dynamic and uses TDMA-based protocol was proposed in [50]. Nodes that have nothing to send or receive during their active cycle go back to sleep immediately to save energy [48]. This is an improvement on protocols like low-energy adaptive clustering hierarchy (LEACH) that remains active during the listen frame even though there are no data to transmit. For nodes with more packets than can be sent within the scheduled duty cycle, cluster heads, dynamically assign different time slots to these nodes in accordance with their needs [49]. Thus protocols are being designed to be more application sensitive and responsive without compromising energy efficient. Energy depletion of the cluster head because of heavy traffic is mitigated by round-robin-based algorithm used for efficient rotation of cluster headship. Energy is saved for nodes with low data traffic while enhancing channel utilisation and for heavy traffic nodes; latency is reduced while there is increase in throughput.

### 3.1.2. Summary of Controlled Access Protocol.

Generally CA protocols conserve nodes energy by avoiding collision since all the nodes are allocated timeslots during which they can transmit. For events that occur at a regular intervals, [47] saves energy by ensuring that node's wake up coincides with event occurrence. But it is not scalable and idle listening addressed in [50] still occurs. Adaptive and dynamic protocols were presented [48] [49] [50]. These adaptively turn ON or off nodes according to the network traffic load demands. This technique saves energy that could have been wasted in idle listening for fixed duty cycles and enhances channel utilisation.

### 3.2. Random Access (RA) Protocol

RA protocols are less compounded than the CA protocols and also they can be completely distributed thus endangering more scalability [51]. CSMA/CA is used by the nodes to access the medium with no master-slave relationships but all nodes compete to gain access to the channel. Less processing and smaller memory are required in RA because of no need to schedule all the nodes thereby reducing control overhead which is the main source of energy drain in CA. Invariably, the rate of collision is higher and actually the main concern in RA protocols. CSMA/CA, though has good scalability, consumes more power and offers low bandwidth utilisation during heavy traffic. Also RA uses preamble sampling or low power listening which occupies the channel for longer time than data packets while hidden stations' preambles keep colliding.

### 3.2.1 Review Random Access Protocols

The probability of collision remains constant with fixed contention windows. This means that all the sensors will compete during each successive contention window after a collision. Collision entails loss of packets and retransmission which waste energy. A technique to enhance the contention window was proposed in [25]. The probability of collision was reduced by halving contending probability which is the likelihood that a sensor will wake-up and contend to access the medium. Hence for any collision, the number of sensors that will be contending in the subsequent contention period will be reduced by half, thus generating a probability sequence $\frac{1}{2}, \frac{1}{4}, \frac{1}{8} \ldots$ and simple probability computation reduces overhead.

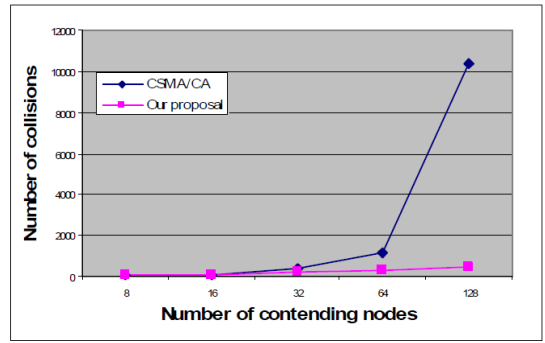

Figure 6: CSMA\CA vs protocol presented in [25]

Figure 6 shows that collisions increase as the number of nodes increase. The protocol proposed in [25] is more energy efficient as it reduces the number of collisions considerably. Less collision preserves energy since there will be less retransmissions. Nonetheless, collisions still occur and many nodes waste energy being awake and contending for channel access. Also in large networks, contending windows will be quite large, wasting energy, increasing latency and lowering bandwidth utilisation. NanoMAC[24] is a non-persistence CSMA/CA energy saving scheme. Contending nodes do not need to continuously listen to the medium, but sleep randomly in the contention window only sensing after backoff. Energy wasted by idle listening is conserved sine nodes listen randomly to the medium. However energy is wasted in carrier sensing and in collision of control packets. More energy is saved by similar protocol, High Efficient Sensor MAC [52] which minimises idle listening by allowing longer sleep periods. Nevertheless idle listening is not completely eliminated since nodes stay awake in case there are packets to transmit. All the nodes hear any on-going transmission and these constitute sources of energy inefficiency. A concern not addressed by [24] and [52] is collision still occurring after the back-off time. Algorithm proposed to reduce collision after back-off time was proposed in [53]. In this algorithm, a node randomly selects back-off period, and notifies others which then select their corresponding back-off time avoiding overlapping. Collision is thus eliminated thereby saving energy but overhead increases which consume more energy.

With reduced overhead and collision, [54] improved on the energy conservation of ML-MAC. The paper proposed a protocol called Adaptive Energy Efficient MAC protocol for Wireless Sensor Networks (AEEMAC). It reduces overhearing by causing nodes with no packets to send to go to sleep upon receipt of Clear-To-Send (CTS) destined for other nodes. There are three optimisation stages in the protocol. The first optimisation tries to reduce idle listening of standard S-MAC protocol by incorporating the duration of the communication in the control packets. When anode overhears the CTS, it knows the duration of the communication and goes to sleep until the end of the communication. The second optimisation considers the actual traffic load of the network and if there are no nodes with packets to send during the active cycle, the nodes will go back to sleep immediately. The last optimisation inserts RTS in ACK packets, reducing overhead and collision while engendering good channel utilisation. But if there are no packet bursts at the beginning of cycle and no multi-hop communications, the protocol will not be effective in saving energy.

### 3.2.2. Summary of Random Access Protocols

RA protocol is more scalable and do not incur much overhead costs, nonetheless, it is prone to collisions. Fixed probability of collision due to rigid contention window was addressed by [25]. It reduced the probability of contention by half whenever collision occurs while [24] uses probability, *p-1,* to refrain from contending for the medium. In both instances, energy was conserved by reducing collision though overhead was incurred. To enhance energy efficiency [52] reduced idle listening but suffer from repeated collisions after backoff. Collisions occurring even after backoff timer in [24] and [52] were addressed by [53]. None interfering backoff timer was employed to improve on the above protocols. Energy was saved as collisions were eliminated but control overhead increased. Dynamic combinations of control packets were used by [54] to reduce overhead thus increasing energy conservation [53]. Further energy preservation was achieved by making nodes that are not involved in transmission to go into energy saving sleep mode thus reducing idle listening.

## 3.3. Slotted Protocol (SP)

SP are used to avoid partial collision which is as a result of packets colliding with a part of another packet. Partial collision has the same effect as full collision as all the packets are lost. However, with SP, frames are divided into slots with duration longer than that required for a packet transmission. Stations are allowed to transmit only at the start of each slot hence collisions can only occur at the beginning of slot. The vulnerable time is equal to one frame duration [41].

### 3.3.1. Review of Slotted Protocol

A Self-Reorganising Slot Allocation protocol was proposed in [55] in which a TDMA MAC frame is kept by each cluster independent of other clusters. Inter cluster collision is avoided by carrier sensing. Whenever the medium is sensed busy, a *carrier sense-collision* is declared which informs the cluster head that there is an overlapping slot. This and hidden terminal issue were resolved by cluster heads reorganising slot allocations after each TDMA frame whenever any of these occurs. This approach increases overhead and there is also the probability that two cluster heads will embark on reorganisation and end up with overlapping slots all the time since slots are assigned independently. Frame scaling includes empty slots and increases frame duration while time slots are allocated to nodes irrespective of whether they have data to send or not. These reduce spectrum efficiency and increase latency. Most TDMA protocols assume that event detection is deterministic but in reality this is not so hence the need for an adaptive protocol that will take into consideration the non-deterministic nature of events. Dynamic Slot Assignment protocol was proposed in [56] and [57], to minimise the effect of nodes occupying the channel when they have no data to transmit. Cluster heads allocate time slots dynamically only to nodes with packets to transmit to cover transmission of all the packets. This saves idle energy and improves bandwidth employment. Nevertheless, if network traffic is heavy, the number of slot requests may exceed the available slots leading to loss of packets.

### 3.3.2. Summary of Slotted Protocols

In [55] cluster heads assign slots using TDMA to avoid collision and save energy but inter-cluster collisions might occur as a result of coinciding slots. Slot allocation in this protocol is fixed and universal leading to slot allocation to nodes without data to send. A dynamic protocol was introduced in [56] and [57] which dynamically assign slots to only nodes with data to send thereby improving on slot allocation of [55]. This eliminates the energy wastage, and improves throughput and bandwidth efficiency.

### 3.4. Hybrid Protocols (HP)

In low traffic, TDMA of CA protocol offers low channel utilisation, while in heavy traffic, CSMA of RA protocol is beset with collisions. Hybrid protocols were developed to combine the advantages of the CSMA, TDMA and other energy efficient MAC protocols to maximise energy efficiency, improve latency and spectrum utilisation.

### 3.4.1. Review Hybrid Protocols

Since no MAC protocol addressed all the sources of energy inefficiencies, hybrids of some of them were introduced to maximise the benefits of various protocols. Zebra-MAC (Z-MAC) [58] was developed to mitigate the short comings of the CSMA and TDMA based protocols while harnessing their advantages. Nodes perform carrier sensing prior to accessing the medium but priority is always given to nodes that own the slot. Each node is assigned a time slot but if it does not have any data to send other nodes will contend for the channel after a predefined set time. Only the slot owner and its one-hop neighbours can contend for the medium in high contention level (HCL). But all nodes can contend in the low contention level (LCL). Explicit congestion notification (ECN) messages are broadcast by a node, upon sensing heavy traffic on the network, to its two-hop neighbourhood to avoid hidden terminal problem. The protocol dynamically uses CSMA and TDMA in light and heavy traffics respectively. Since CSMA is more energy efficient in low traffic by avoiding idle listening, the protocol saves energy. The use of TDMA in heavy traffic reduces collision hence energy is preserved, thus engendering high channel utilisation. Idle listening, waiting for set time to elapse and clear channel assessment (CCA) all contribute to energy depletion and low throughput. Synchronisations of nodes within two hops and switching between TDMA and CSMA have overhead cost also.

A centralised hybrid scheme that uses both the principle of modified slotted contention-based and contention free protocols to preserve energy, was proposed in [59]. It improved on the energy consumption of BMA [60] by reducing control overhead. The cluster heads broadcast a schedule for all nodes with data to send while nodes without data go to sleep [56] and [57]. Synchronisation and defining of the superframe structure were done with beacons to reduce overhead and save energy. No node is required to know its ordering number or to synchronise with its one-hop or two-hop neighbours like in Z-MAC [58]. Compliance period and reservation were intruded to reduce overhearing and collision thereby conserving energy, while dynamic slot allocation improves channel efficiency. However, the protocol dealt only with intra-cluster collisions, a protocol that dealt with both intra [59] and inter-cluster collision was proffered in [61]. CA based TDMA, which reduces energy waste due to collision was used for intra-cluster medium access. RA based CSMA was used for inter-cluster spectrum access among cluster heads thus reducing energy inefficiency of control overhead. However in order to be responsive to network demands, adaptive sleep/active cycle was employed. Nodes on active mode if they have no packets to send or receive go back to sleep immediately, whereas those with more packets have their active cycle increased. Multi-hop communication, which improves energy efficiency, was used in [61] for transmission from cluster heads to the sink. But cluster heads are permanently on active mode and this will lead to energy wastage. Since TDMA is used in intra-cluster communications, during low traffic, there will be low channel utilisation. On the other hand, during heavy inter cluster communications, collisions will increase.

An emergency response (ER-MAC) hybrid protocol that works on a similar principle as Z-MAC, but saves more energy by avoiding contention by a node that owns a slot, was developed in [62]. Also it improves on [61] by eliminating permanently on cluster heads thus saving more energy. It is a multi-hop tree protocol that can be applied in events like patient monitoring, wild fire and intruder detections. In these applications, there may be no activity for a long time and suddenly, there might be an event sensed by different nodes that might require immediate reporting, thus bursts of packets. Packets are queued and prioritised in the protocol, and high priority packets are transmitted before low priority packets.

Though slot owner with high priority packet are exempted from contention, maintaining an update of the time remaining before a packet deadline expires implies lots of overhead control cost. The use of fixed frame in this protocol means that, it is not adaptive to the dynamics of network load traffic variations. Queue-length aware MAC (Queue-MAC) [63] is a multi-hop beacon enabled hybrid MAC protocol that addressed the issue of fixed cycle of [62]. It incorporates a dynamic duty cycled TDMA while the CSMA duty cycle remains fixed. This allows frames to be dynamically adjusted to make room for the transmission of more packets within a frame. Similarly, CSMA and TDMA are used interchangeably according the volume of traffic. Accordingly, making the protocol suitable for applications with fluctuating traffic and saving energy that would have been wasted for idle listening and collisions. Nonetheless, beacon, ACK packets and updating of the queue length indicator table will lead to increase in overhead energy cist.

### 3.4.2. Summary of Hybrid Protocols

Hybrid techniques were developed to optimise the gains of different protocols which are combined into one single protocol. Z-MAC [58] uses a combination of CSMA and TDMA to enhance energy efficiency, greater spectrum utilisation and throughput. But it involves synchronisation within two-hop neighbourhood, CCA and idle listening which are all sources of energy wastages. An improvement on the overhead cost of [58] was made in [59] by using beacons for synchronisation and specifying superframe structure thus reducing overhead. Nevertheless, exchange control packets causes collisions and low spectrum usage, while long direct communication with the sink wastes energy. Inter and intra cluster link was the focus of [61], using TDMA for intra cluster link, and CSMA for inter-cluster communication which also improved on [59]. But having cluster heads permanently on the active mode wastes energy. Contention by slot owner with priority packets in Z-MAC was eliminated [62]. Nonetheless fixed frame of the scheme means that nodes with more packets than can be sent in one superframe cannot finish the transmission in one cycle. This was addressed in Queue-MAC [63] by the introduction of dynamic superframe with variable TDMA slot. The flexibility allows nodes to send all their packets in one superframe since the superframe can be adjusted according to the load demands of nodes. Nonetheless, overhead is incurred by the use of beacons and ACK.

Inroad has been made in developing energy efficient MAC protocols but challenges still abound as we have highlighted in this survey. Collision in RA when traffic is heavy has not been comprehensively addressed. Though dynamic cycle is improving on the energy wasted in idle listening in CA protocols under light traffic it requires nodes to access the medium for a short time in case it has packets to transmit.

## 4. CHALLENGES

Sources of energy inefficiencies in WSN have been highlighted in section 2 and a detailed review of WSN MAC protocols is presented in section 3. The challenges throw up from the survey are outlined in this section. Source of constant power supply has remained a major challenge in the development of WSNs. Suitable energy harvesting algorithm has not been developed and a steady or constant source of power supply for WSNs will significantly enhance the network. There have been improvements on channel utilisation and energy preservation by adaptive and dynamic protocols but the overhead is still high. Cross channel communication is an issue with channelization protocols as nodes have to switch between the channels. Switching between protocols is also a challenge in the implementation of hybrid protocols. Contention based protocols though provides good scalable network with efficient bandwidth utilisation, it wastes energy under heavy traffic and leads to network degradation. On the other hand control packets in CA waste energy. Complete eradication of these can transform WSN and its application. Death of nodes means that a network wide scalability should be introduced.

Routing consumes almost same amount of energy as communication. Routing topologies often make use of relay nodes like parent nodes or cluster heads. This causes uneven energy depletion on nodes at the root level or cluster heads.Another challenge is continuous sensing of same event. Some events do not require continuous sensing since the outcome does not change constantly. Data aggregation is being employed to reduce redundancy sampling but this may require nodes to perform data aggregation before forwarding packets. Possibility of node failure is a challenge as this will cause loss of all previous data from the source node en route to the sink. Also many nodes may sense the same event and take different routing parts to forward the data, which means that even data aggregation may not stop the transmission. Unneeded sampling requires further investigation as the sensing wastes both routing and communication energy. One other source issue is the energy used for maintaining the routing table as nodes need to know the next node nearest to the sink, reducing this will enhance network lifetime.

## 5. CONCLUSIONS AND FUTURE WORK

This paper reviewed MAC protocol for WSNs, sources and causes of energy inefficiencies and their consequences on the network. WSNs demand tailored MAC protocols because of energy constrains which is not as much an issue with many other wireless networks. Network lifetime extension through energy proficient MAC protocols is reviewed by grouping MAC protocols into four categories according to the methodology of medium access deployed. From the available literature, MAC protocols have evolved from fixed duty cycle or contention windows to adaptive, dynamic and flexible protocols. This has greatly enhanced energy efficiency especially when different protocols are merged in hybrid algorithms.

Most protocols involving clusters use direct communication link between the cluster head and the sink. Research into using a more energy efficient multi-hop communication protocol between the cluster heads and the sink will help conserve the energy, and accordingly extending the lifetime of the WSN. Most of the research on energy efficient controlled access protocol has come at a cost of control packets overhead. By investigating this research area further, an enhanced energy efficiency protocol may be developed that can revolutionise WSN's power consumption. Another challenge with CA protocols is that they are not scalable and this can be investigated further. Though CDMA involves complex computing, but also it can eliminate collision, developing a technique for wide application of this in WSN will boost energy efficiency, and hence, network lifetime of WSNs. This will be more so since processing consumes less energy than communication. Like CDMA, FDMA might improve on the energy budget of WSNs with appropriate protocol. TDMA appeared to be the leading base protocol for energy efficiency in WSNs

**Authors**

Eleazar Chukwuka obtained HND in Electrical/Electronic Eng in 2001, BEng and MSc from University of Greenwich, in 2009 and 2012 respectively. His research interests are in sensor networks and cognitive radio.

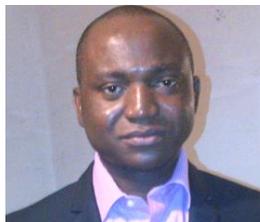

Dr Kamran Arshad received his BEng, MSc and PhD in Electrical Engineering with Distinction in 2000, 2003 and 2007 and currently working as a Senior Lecturer at University of Greenwich. Dr Arshad research interests are in general area of cognitive radio, LTE-Advanced and in general area of future networks. He is an author of a book titled Radio wave propagation modelling using finite element method (Lambert Publishing, 2010) and author of over 80 technical papers published in peer-reviewed international Journals and conferences. He received 03 best paper awards and chaired 03 technical sessions in leading international conferences. He served more than 30 conferences as a Technical Programme Committee (TPC) member and editor of International Journal of Artificial Intelligence.

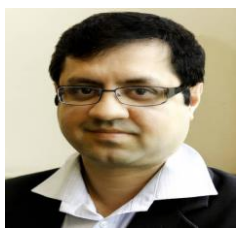

Dr Arshad was officially invited to become a panellist in IEEE PIMRC 2011 workshop on cognitive radio, Toronto, Canada (September 2011). He was also invited as a speaker in International Software Radio Symposium held in London, UK (June 2012).